\documentclass[aps,prl,twocolumn,showpacs,amssymb, footinbib]{revtex4-1}
\usepackage{amsfonts,amsmath,amssymb,amsthm,bbm,hyperref}
\usepackage{color,psfrag}
\usepackage[T1]{fontenc}
\usepackage[dvips]{graphicx}

\usepackage[normalem]{ulem}

\newcommand{\ket}[1]{\left| #1 \right\rangle}

\newcommand{\braket}[2]{\left\langle \vphantom {#1 #2} #1 \hphantom{|} \right| \left. \vphantom {#1 #2} #2 \right\rangle}

\newcommand{\be}{\begin{equation}}
\newcommand{\ee}{\end{equation}}
\newcommand{\ba}{\begin{eqnarray}}
\newcommand{\ea}{\end{eqnarray}}

\begin{document}

\title{A quantum reduction to Bianchi I models in loop quantum gravity}

\author{N. Bodendorfer}
\email[]{norbert.bodendorfer@fuw.edu.pl}

\affiliation{Faculty of Physics, University of Warsaw, Pasteura 5, 02-093, Warsaw, Poland}

\date{{\small \today}}

\begin{abstract}

We propose a quantum symmetry reduction of loop quantum gravity to Bianchi I spacetimes. To this end, we choose the diagonal metric gauge for the spatial diffeomorphism constraint at the classical level, leading to an $\mathbb{R}_{\text{Bohr}}$ gauge theory, and quantise the resulting theory via loop quantum gravity methods. Constraints which lead classically to a suitable reduction are imposed at the quantum level.
The dynamics of the resulting model turn out to be very simple and manifestly coincide with those of a polymer quantisation of a Bianchi I model for the simplest choice of full theory quantum states compatible with the Bianchi I reduction. In particular, the ``improved'' $\bar{\mu}$ dynamics of loop quantum cosmology can be obtained by modifying the regularisation of the Hamiltonian constraint with similar ideas, in turn yielding insights into the full theory dynamics. 
\end{abstract}


\maketitle

\section{Introduction}
Identifying symmetry reduced sectors within full theories is an important problem, since success in this endeavour usually allows one to perform computations which are otherwise intractable. Within loop quantum gravity (LQG), there has been a lot of recent interest in this subject, see e.g. \cite{BojowaldSymmetryReductionFor, BojowaldSphericallySymmetricQuantum, EngleRelatingLoopQuantum, BrunnemannSymmetryReductionOf, BianchiTowardsSpinfoamCosmology, AlesciANewPerspective, EngleEmbeddingLoopQuantum, GielenCosmologyFromGroup, HanuschInvariantConnectionsIn}. Different strategies can be employed towards this goal, the most active one being the identification of suitable symmetry reduced states directly within the full, not classically gauge fixed, quantum theory. While success along this route might be preferable, we will deal with another approach in this paper, which consists of choosing a gauge fixing at the classical level, adapted to the symmetry reduction that one wants to achieve. In particular, we are going to gauge fix the spatial metric to be diagonal (also called an orthogonal system \cite{DeTruckExistenceOfElastic}), which is a gauge fixing admitted by Bianchi I models. Our strategy will then be to quantise this model and impose a symmetry reduction at the quantum level. This strategy clearly separates the steps of gauge fixing and symmetry reduction, which is less transparent when performing both at the quantum level. A related proposal is \cite{AlesciANewPerspective}, which inspired us to the present paper in the first place. The approach taken in this paper is similar to the one in \cite{BLSI}, where a reduction to spherical symmetry is achieved within a quantisation of general relativity in the radial gauge.

\section{The diagonal metric gauge}
We start with the ADM formulation of general relativity, that is with the phase space coordinatised by the spatial metric $q_{ab}$ and its momentum $P^{ab}$, living on the spatial slice $\Sigma$ of $3$-torus topology, with the canonical Poisson brackets $\left\{q_{ab}(\sigma ), P^{cd}(\sigma') \right\} = \delta_{(a}^c \delta_{b)}^d \delta^{(3)}(\sigma,\sigma')$, subject to the Hamiltonian constraint $H$ and the spatial diffeomorphism constraint $C_a := -2 \nabla_{b} P^b {}_a = 0$. We now introduce the gauge fixing $q_{ab} = 0$ for $a \neq b$, i.e. $q_{ab} = \text{diag}(q_{xx}, q_{yy}, q_{zz})_{ab}$, for the spatial diffeomorphism constraint, which is at least locally accessible \cite{DeTruckExistenceOfElastic}. We note that not all spatial diffeomorphisms are gauge fixed by this condition, but only those which do not preserve the off-diagonal components of $q_{ab}$. In particular, $C_a$ smeared with a lapse function of the form $\vec N = (N^x(x), N^y(y), N^z(z))$ is still a first class constraint \footnote{Due to the $3$-torus topology, there are no fall-off conditions at spatial infinity prohibiting such shift vectors, see \cite{AlbaTheEinsteinMaxwell1, AlbaTheEinsteinMaxwell2, AlbaTheEinsteinMaxwell3} for the asymptotically flat case.}. We will call diffeomorphisms generated by such shift vectors ``restricted''.

We would now like to go to the reduced phase space, which is coordinatised by $q_{xx}, q_{yy}, q_{zz}$ and $P^{xx}, P^{yy}, P^{zz}$, and obtained by solving the second class pair $q_{a \neq b} = 0 = C_a$.
While solving $q_{a \neq b} = 0$ is straight forward, we need to compute an expression for $P^{xy}, P^{xz}$, and $P^{yz}$ in terms of the reduced phase space coordinates by using $C_a = 0$. For, say, $P^{xy}$, this is, due to linearity of $C_a$ in $P^{ab}$, equivalent \footnote{Equation (\ref{eq:PxyReduced}) is the only non-trivial contribution from a gauge unfixing projector \cite{VytheeswaranGaugeUnfixingIn} acting on $P^{xy}$.} to solving the equation 
\be
	\left. \frac{\delta}{\delta P^{c \neq d}(x)} \left( P^{xy}[\omega_{xy}] + C_a [N^a]\right)  \right|_{q_{a \neq b}=0} = 0  \label{eq:PxyReduced}
\ee
for $N^a$ as a functional of $\omega_{xy}$, where by $P^{xy}[\omega_{xy}]$ we mean the smearing $\int_\Sigma d^3\sigma P^{xy}(\sigma) \omega_{xy}(\sigma)$. Given $N^a$ as a function of $\omega_{xy}$, we can evaluate $\tilde P^{xy}[\omega_{xy}] := P^{xy}[\omega_{xy}] + C_a [N^a]$ at $P^{c \neq d} = 0 = q_{c \neq d}$ since $\tilde P^{xy}$ does not depend any more on $P^{c \neq d}$ for $N^a$ solving \eqref{eq:PxyReduced}, and obtain the desired expression for $P^{xy}$ on the reduced phase space, i.e. as a functional of $q_{xx}, q_{yy}, q_{zz}$ and $P^{xx}, P^{yy}, P^{zz}$. By construction, $\tilde P^{xy}$ preserves the gauge fixing condition $q_{a \neq b} = 0$.
Up to a boundary term, we have 
\begin{align}
	&\tilde P^{xy}[\omega_{xy}] \label{eq:GenSpatialQP} \\
	= &\left. \int_\Sigma d^3 \sigma \left(P^{xx} \mathcal L_{\vec{N}} q_{xx}+P^{yy} \mathcal L_{\vec{N}} q_{yy}+P^{zz} \mathcal L_{\vec{N}} q_{zz}\right) \right|_{q_{a \neq b}=0}  \text{,} \nonumber 
\end{align}
where $\mathcal L_{\vec N}$ denotes the Lie derivative with respect to the vector field $\vec N$. 

For a general smearing $P^{ab}[\omega_{ab}]$, 
we obtain from \eqref{eq:PxyReduced} the equations
\begin{align}
	2 \nabla_{(x} N_{y)} &= q_{yy} \partial_x N^y + q_{xx} \partial_y N^x =  \omega_{xy} \label{eq:X}\\ 
	2 \nabla_{(y} N_{z)} &= q_{zz} \partial_y N^z + q_{yy} \partial_z N^y =  \omega_{yz} \label{eq:Y} \\
	2 \nabla_{(z} N_{x)} &= q_{xx} \partial_z N^x + q_{zz} \partial_x N^z =  \omega_{zx} \label{eq:Z} \text{.}
\end{align}
A general solution to these equations might be hard to find, however it is not needed for what follows. Instead, we will show that by choosing $\omega_{a\neq b}$ appropriately, we can generate arbitrary vector fields $N^a$.
In particular, choosing $\omega_{yz} = 0$, we find the solution $N^y = 0 = N^z$, and 
\be
	N^x(x,y,z) =  \int^y d y' \, q^{xx}  (x,y',z) \omega_{xy} (x,y',z) \text{,} \label{eq:Nx}
\ee
along with the consistency condition
\be
	N^x(x,y,z) =  \int^z d z' \, q^{xx}  (x,y,z') \omega_{xz} (x,y,z') \text{.} \label{eq:Consistency}
\ee
We can now choose $\omega_{xy}$ to generate an arbitrary $N^x$ in \eqref{eq:Nx}, while choosing $\omega_{xz}$ to satisfy \eqref{eq:Consistency} \footnote{For non-trivial topology of $\Sigma$, one encounters topological obstructions for $\vec N$ to be globally well defined. We will restrict us to such $\omega$ that these conditions are met.}. The argument can be repeated to generate an arbitrary $N^y$ and $N^z$ by adding the respective $\omega_{a \neq b}$. 
We could choose to exclude diffeomorphisms of the restricted type by properly adding the corresponding spatial diffeomorphism constraints, however this is not of importance, as we will see later. 
We conclude that the complete set of $P^{a \neq b}[\omega_{a \neq b}]$ on the reduced phase space, along with the restricted spatial diffeomorphisms, corresponds to a complete set of generators of spatial diffeomorphisms acting on the reduced phase space if $P^{a \neq b}=0$ is satisfied. This last condition is necessary in order for \eqref{eq:GenSpatialQP} to generate spatial diffeomorphisms in both $q_{aa}$ and $P^{bb}$. The condition $P^{a \neq b}[\omega_{a \neq b}] = 0$ thus corresponds to implementing spatial diffeomorphism invariance for arbitrary shift vector fields on the reduced phase space, {\it on top} of having solved large parts of the spatial diffeomorphism constraint already via a gauge fixing. 

Our strategy will now be to kinematically quantise the above system, which still corresponds to full general relativity. We will not define the full theory Hamiltonian constraint, since this requires the solution of the equations \eqref{eq:X}, \eqref{eq:Y}, and \eqref{eq:Z}. Instead, we will impose $P^{a \neq b} = 0$ as invariance under finite spatial diffeomorphisms along with another constraint, which will lead to a quantum system capturing the degrees of freedom of a Bianchi I model, yet leaving room for inhomogeneities. On the corresponding quantum states, one can quantise the classical Hamiltonian evaluated at $P^{a \neq b} = 0$, which is consistent with the reduction.
Since $P^{a \neq b}[\omega_{a \neq b}]=0$ is an additional condition on the reduced phase space, the passage to spatially diffeomorphism invariant states has to be interpreted as a reduction of {\it physical} degrees of freedom.\\

\section{Quantum theory}
We first need a set of variables suited for a LQG type quantisation. We define $e_a := \sqrt{q_{aa}}$ with no summation implied. $E^a := \sqrt{\det q} \, e^a$ corresponds to the densitised triad of the Ashtekar-Barbero variables. Next, we define $K_x := K_{xx} e^x$, where $K_{ab}$ is the extrinsic curvature, appearing in $P^{ab} = \frac{\sqrt{\det q}}{2} \left( K^{ab} - q^{ab} K \right)$. The new Poisson brackets read
\be
	\left\{ K_a(\sigma), E^b(\sigma') \right\} = \delta_a^{b} \delta^{(3)} (\sigma,\sigma') \text{,} \label{eq:NewPoissonBrackets}
\ee
and the spatial diffeomorphism constraint (for arbitrary shift vector) becomes, up to a boundary term, 
\be
	C_a[N^a] = \int_\Sigma d^3 \, \sigma E^a \mathcal L_{\vec N} K_a + \ldots  ~~ \text{.} \label{eq:DiffConstraint}
\ee
$\ldots$ stands for terms of the form $\partial_a e_b$, $a \neq b$, and $\partial_a K_b$, $a \neq b$, which vanish for the special case of restricted diffeomorphisms, and originate from the fact that $\mathcal{L}_{\vec N} q_{ab} \neq 2 e_{(a} \mathcal{L}_{\vec N} e_{b)}$, even for $a=b$. Up to these terms, $K_a$ transforms as a one-form, and $E^a$ as a densitised vector. Since we would like to have an explicit interpretation of the action of $C_a$ as spatial diffeomorphisms, we will impose the additional constraints 
\be
	\partial_a e_b = 0 = \partial_a K_b, ~~a \neq b \text{,} \label{eq:AdditionalConstraints}
\ee
up to which \eqref{eq:DiffConstraint} generates spatial diffeomorphisms. We note that these constraints are fully consistent with the Bianchi I symmetry. Moreover, they are first class with respect to the restricted diffeomorphisms. In fact, they e.g. impose $e_x = e_x(x)$, such that the remaining $x$-dependence is removed by the restricted diffeomorphisms in $x$-direction. The proper generator of spatial diffeomorphisms, i.e. the one without the $\ldots$, will be denoted by $\bar C_a$.

Since we want to perform a Dirac-type quantisation, i.e. impose constraint operators on the kinematical Hilbert space, we need to pick a first class subset of \eqref{eq:DiffConstraint} and \eqref{eq:AdditionalConstraints}, also known as ``gauge unfixing'' the above constraint system \cite{MitraGaugeInvariantReformulationAnomalous,  VytheeswaranGaugeUnfixingIn}. First, we choose $\bar{C}_a = 0$, since it can be implemented in the quantum theory by the methods developed in \cite{AshtekarQuantizationOfDiffeomorphism}. Next, we need to pick further constraints from \eqref{eq:AdditionalConstraints}, which (upon quantisation) are first class with $\bar{C}_a = 0$ as well as the Hamiltonian constraint. A natural choice is $G := \partial_a E^a = 0$, the Gau{\ss} law, which we also adopt due to its simplicity. While not having a rigorous proof, we do not believe that additional first class constraints with the above properties can be found within \eqref{eq:DiffConstraint} and \eqref{eq:AdditionalConstraints}. In any case, imposing $\bar{C}_a = 0 = G$ in the quantum theory turns out to be sufficient to reduce the quantum states in such a way that the simplest reduced states exactly encode the degrees of freedom of a Bianchi I model and that the dynamics agrees with the corresponding minisuperspace quantisation. 

Let us now deal with the kinematical quantisation, based on \eqref{eq:NewPoissonBrackets}. At this stage, the theory still encodes all degrees of freedom of general relativity (if compatible with the diagonal gauge).
We will treat $K_a$ in analogy to the Ashtekar-Barbero connection in LQG: we define the holonomies $h^\lambda_\gamma(K) := \exp \left( i \lambda \int_\gamma  K_a \,d s^a \right) $
for an arbitrary oriented path $\gamma$ and $\lambda \in \mathbb{R}$ labelling a representation of the group $(\mathbb{R}, +)$. By smearing $E^a$ over two-surfaces, we obtain fluxes, and consequently a standard holonomy-flux algebra of an $\mathbb{R}$ gauge theory. Holonomies can be readily generalised to charge networks, the Abelian analogues of spin networks. 
It is also possible to introduce a Barbero-Immirzi-like parameter by rescaling $E^a, K_a$ accordingly. A similar system, Maxwell theory, has been quantised by the same methods in \cite{CorichiAmbiguitiesInLoop}.

Quantisation can now be achieved via the Gelfand-Neimark-Segal construction by specifying the positive linear Ashtekar-Lewandowski functional $\omega_{\text{AL}}$ on the holonomy-flux algebra \cite{AshtekarRepresentationsOfThe, AshtekarRepresentationTheoryOf}. Since this functional is only well defined for compact groups, we need to substitute $\mathbb{R}$ by its Bohr compactification $\mathbb{R}_{\text{Bohr}}$, which is compact and admits a normalised and translation-invariant (Haar) measure $\mu_{\text{H}}$, given by (see e.g. \cite{SubinDifferentialAndPseudodifferential})
\be
	\mu_{\text{H}}(f)  = \int_{\mathbb R_\text{Bohr}} d \mu_{\text{H}} \, f(x):= \lim_{R \rightarrow \infty} \frac{1}{2R} \int_{-R}^R dx \, f(x) \text{.}
\ee
An inner product can now be defined as
\be
	\braket{h^\lambda_\gamma}{h^{\lambda'}_\gamma} :=  \int_{\mathbb R_\text{Bohr}} d \mu_{\text{H}} \,\overline{e^{i \lambda x}} e^{i \lambda' x} = \delta_{\lambda, \lambda'} \label{eq:InnerProduct}
\ee
for two holonomies defined on the same edge, and following the usual construction \cite{AshtekarRepresentationTheoryOf} for more complicated graphs.
General cylindrical functions over a single edge can be identified with almost periodic functions \cite{SubinDifferentialAndPseudodifferential}.
Completion with respect to \eqref{eq:InnerProduct} then yields the Hilbert space $L^2({\mathbb R}_{\text{Bohr}}, d \mu_{\text{H}})$ for each edge.
Within loop quantum gravity, this Hilbert space is well known from loop quantum cosmology \cite{AshtekarMathematicalStructureOf}.

We can now define an operator measuring the area $A$ of a surface $S$ by substituting the flux operator in the expression $A(S) = | \int_S E^a \,d^2 s_a  |$. The important difference from the usual SU$(2)$ case \cite{SmolinRecentDevelopmentsIn} is that we do not need to define the area operator as $\int \sqrt{|\text{flux}^2|}$, since $E^a$ is gauge invariant as opposed to the SU$(2)$ densitised triad $E^{a}_i$. Thus, in analogy to electric charge, the area operator of a closed contractible surface (electric flux through $S$) we consider vanishes e.g. on a single contractible Wilson loop, since the contributions coming from two intersections always cancel. 

On this Hilbert space, the $h^\lambda_\gamma(K)$, seen as cylindrical functions, provide a basis and classically separate points on the configuration space. They are further subject only to the Hamiltonian constraint and the restricted set of spatial diffeomorphisms. In particular, these restricted spatial diffeomorphisms are not sufficient to reduce our quantum states to diffeomorphism equivalence classes, since they only contain a small subset of all shift vectors, but the Hilbert space all possible graphs. Thus, after modding out the restricted spatial diffeomorphisms, the quantum states still know about the embedding information of the paths $\gamma$ into $\Sigma$, which is in stark contrast to the spatially diffeomorphism invariant Hilbert space of LQG \cite{AshtekarQuantizationOfDiffeomorphism}.

We are now in a position to discuss a reduction of the proposed quantum theory to Bianchi I models. Following the previous discussion, we implement $\bar{C}_a = 0 = G$ on the kinematical Hilbert space. Gauge invariance, following from $G = 0$, is easily incorporated by only using charge networks satisfying $\sum_i \lambda_i =  0$ at vertices, where the orientations of the edges are chosen to coincide. Spatial diffeomorphism invariance, following from $\bar{C}_a=0$, is implemented using the methods of \cite{AshtekarQuantizationOfDiffeomorphism}, roughly by going over to diffeomorphism equivalence classes of graphs. 
The resulting picture thus mirrors the quantisation of Abelian BF theory, see the seminal work \cite{HusainTopologicalQuantumMechanics}, as well as \cite{HusainBackgroundIndependentDuals}.

Since we restrict the topology of $\Sigma$ to be a $3$-torus $\mathbb T^3$, we can naturally identify three 2-tori $\mathbb T^2_x$, $\mathbb T^2_y$, $\mathbb T^2_z$, orthogonal to the $x$, $y$, and $z$-direction. Wilson loops on $\Sigma$ can be either contractible, or wrap around the three 1-tori $\mathbb T^1_x$, $\mathbb T^1_y$, and $\mathbb T^1_z$, in the $x$, $y$, and $z$-direction.

The most elementary example $\ket{\lambda_x, \lambda_y, \lambda_z}$ of a charge network describing a Bianchi I universe with $\mathbb T^3$ topology and satisfying all the above constraints is to consider three Wilson loops with $\mathbb{R}$-labels $\lambda_x,\lambda_y,\lambda_z$, wrapping around $\mathbb T^1_x$, $\mathbb T^1_y$, and $\mathbb T^1_z$ respectively with winding number $1$. 
We furthermore require that these Wilson loops intersect in a single $6$-valent vertex. 
We choose three closed surfaces $S_x$, $S_y$, and $S_z$ wrapping around the 2-tori and consider the area operators $\hat A(S_x), \hat A(S_y), \hat A(S_z)$. They are observables with respect to $G$ and $\bar{C}_a$, since their action on gauge invariant charge network states is equivalent for two surfaces $S$ and $S'$ which differ by a spatial diffeomorphism. This property directly results from the Abelian gauge group. 
More precisely, given a Wilson loop along $\mathbb T^1_x$ with winding number $n \in \mathbb Z$ and representation label $\lambda$, $\hat A(S_x)$ acts by multiplying $|n \lambda |$, independently of the local intersection characteristics.
Thus, $\hat A(S_x), \hat A(S_y), \hat A(S_z)$ provide us with (global) observables which we can directly relate to the minisuperspace variables $\tilde E^b = \frac{1}{2}\int_{\mathbb T^2_b} E^c \epsilon_{cde} \, dx^d \wedge dx^e$. Furthermore, on $\ket{\lambda_x, \lambda_y, \lambda_z}$, we define the holonomy observables $h^\lambda_\gamma(K)$ for $\gamma$ coinciding with one of the three non-contractible Wilson loops defining the quantum state.

More complicated states can be constructed by adding Wilson loops, either wrapping around a torus, or contractible, thus changing the local properties of the charge network, e.g. adding new edges and changing the representation labels on existing edges. While contractible Wilson loops do not change the ``homogeneous modes'' $\hat A(S_x), \hat A(S_y), \hat A(S_z)$, they behave as ``inhomogeneous perturbations'' on the homogeneous background, since their presence can be detected by local operators such as the Ricci scalar (to be discussed below). Thus, the present framework naturally incorporates a notion of inhomogeneous perturbations on top of the Bianchi I background. 

The Hamiltonian constraint now needs to be regulated in terms of holonomies and fluxes. As remarked before, we do not know the solution to \eqref{eq:X}, \eqref{eq:Y}, and \eqref{eq:Z}, prohibiting the construction of the full theory Hamiltonian constraint. However, on the space of spatially diffeomorphism invariant distributions, we have $P^{a \neq b}[\omega_{a \neq b}] = 0$, which suggests that one should just quantise the classical Hamiltonian with $P^{a \neq b} = 0$, such that it is consistent with all the constraints imposed. Under this condition, and in the diagonal metric gauge, the Hamiltonian constraint becomes $H = (e_x K_y K_z + e_y K_x K_z + e_z K_x K_y)/2 +   \sqrt{q} R$.
In order to quantise it, we are going to adapt the methods developed in \cite{ThiemannQSD1, GieselAQG4} to our case, focussing on states of the type $\ket{\lambda_x, \lambda_y, \lambda_z}$. 
Thiemann's trick e.g. amounts to $e_x = 2 \{ K_x, V\}$, where $V$ can be taken to be the total volume of the universe. Thus, 
\be
	 H =  \{ K_x K_y K_z, V\} + \sqrt{q} R \text{.} \label{eq:HamSimple}
\ee
The volume operator can be regularised in the standard way following \cite{AshtekarDifferentialGeometryOn}, yielding a diagonal operator due to the Abelian gauge group.
$K_x$ has to be approximated in $H$ via holonomies. The naive choice is simply $\int_{\gamma_x} K_x \, dx \approx \frac{1}{2i} \left(h^1_{\gamma_x}(K)-h^1_{\gamma_x}(-K)\right)$, where $\gamma_x$ is some path in the $x$-direction, and similar for $y$, $z$. However, since the paths $\gamma$ span the whole universe in the state $\ket{\lambda_x, \lambda_y, \lambda_z}$, and we work within a graph-preserving regularisation, thus having $\gamma_x$ coinciding with the corresponding path defining $\ket{\lambda_x, \lambda_y, \lambda_z}$, the approximation will be unsuitable in most situations since $\int_{\gamma_x} K_x \, dx$ will be large in general. In fact, proceeding with this regularisation, we end up with the so called $\mu_0$-dynamics of loop quantum cosmology (LQC) \cite{AshtekarMathematicalStructureOf}, which have been shown to lead to physically unacceptable results \cite{AshtekarQuantumNatureOfAnalytical}. More precisely, we can identify the (isotropic) variables $c,p$ from \cite{AshtekarMathematicalStructureOf} as $c \sim \int_{\gamma_x} K_x \, dx$ and $p \sim \tilde E^x$ (or $y$, $z$).

This problem can be solved by applying the ideas of the ``improved'' $\bar{\mu}$-dynamics of LQC \cite{AshtekarQuantumNatureOf} also in the full theory: instead of using the representation label $\lambda^0_x = 1$, we use $\bar {\lambda}_x = 1 /  \int_{\gamma_x} e_x \, dx$, i.e. we normalise the integrated connection by the appropriate sizes of the universe as $\int_{\gamma_x} K_x \, dx \approx \frac{1}{2i \bar \lambda_x} \left(h^{\bar \lambda_x}_{\gamma_x}(K)-h^{\bar \lambda_x}_{\gamma_x}(-K)\right)$. At the level of dynamics, it follows that departures from classical general relativity only occur in regimes where the Planck density is approached \cite{AshtekarLoopQuantumCosmologyBianchi}. We regularise $\int_{\gamma_x} e_x \, dx$ as an operator as in \cite{AshtekarLoopQuantumCosmologyBianchi} as $\sqrt{|\tilde E^y \tilde E^z / \tilde E^x|}$, and the Hamiltonian has to be symmetrised, e.g. as in \cite{AshtekarLoopQuantumCosmologyBianchi}. 

While we may choose to exclude the Ricci curvature from the Hamiltonian by gauge unfixing arguments (we additionally introduce $\partial_a e_a = 0$, which doesn't change the result of the gauge unfixing), it is still instructive to see that a standard regularisation in terms of fluxes would vanish. This follows from the fact that derivatives of the type $\partial_a E^b$ would be regularised via a neighbouring vertex finite difference approximation, however on $\ket{\lambda_x, \lambda_y, \lambda_z}$, the resulting operator would vanish since the single vertex in the underlying graph is its own neighbour. A similar result would hold for a more refined, yet still purely homogeneous state due to $E^b$ acting diagonally. However, when introducing inhomogeneities via contractible Wilson loops, the situation is expected to change. The precise regularisation of the Hamiltonian constraint for more general quantum states, e.g. along the lines of \cite{GieselAQG4}, will be reported elsewhere.

The above discussion now allows us to conclude that we have achieved a dynamical equivalence between a LQC type quantisation and our computation within a reduced sector of the full theory. The crucial last step is to identify the minisuperspace variables as observables with respect to the constraints responsible for the reduction. In particular, the state $\ket{\lambda_x, \lambda_y, \lambda_z}$ is just mapped into its LQC analogue \cite{AshtekarLoopQuantumCosmologyBianchi} $\ket{p_1, p_2, p_3}$.
The Hamiltonian constraint in the full theory then acts on $\ket{\lambda_x, \lambda_y, \lambda_z}$ in the same way as in LQC \footnote{In \cite{AshtekarLoopQuantumCosmologyBianchi}, the Thiemann trick for $e_x$ is not used, but instead a regularisation via fluxes. We could also adopt this strategy in our case. As soon as both regularisation strategies coincide, equivalence of the dynamics follows. The framework is thus insensitive to such details.}. 
The matter clocks that are usually coupled in LQC can also be incorporated in the full theory by standard means.

\section{Comments} 
From standard SU$(2)$-based LQG, one might expect that the gauge group in the reduced setting would be U$(1)$ \footnote{This was in fact the approach taken in the first preprint version of this paper.}. However, this leads to the $\mu_0$ dynamics of LQC when using a maximally coarse state, and one needs to incorporate also non-integer representations through the use of $\mathbb R_{\text{Bohr}}$ as a gauge group in order to support the $\bar \mu$ dynamics. The discrepancy between the LQC and LQG volume spectra is thus naturally resolved in our framework by using $\mathbb R_{\text{Bohr}}$ instead of U$(1)$ as a gauge group. 
Still, it would be more desirable to see the $\bar \mu$ dynamics emerge from a coarse graining limit of standard SU$(2)$-based LQG, see e.g. \cite{AlesciLoopQuantumCosmology} for recent work.\\
\indent In turn, we can conclude for the full theory that a graph preserving Hamiltonian constraint operator, at least when acting on very coarse states encoding large geometries, should be modified according to the $\bar \mu$ prescription in order to avoid the shortcomings pointed out in \cite{AshtekarQuantumNatureOfAnalytical}. Also here, it would be desirable to see a similar property emerge from a coarse graining limit.

\section{Conclusion}
We have presented a derivation of a Bianchi I sub-sector of a quantisation of general relativity in the diagonal metric gauge, using quantisation methods of LQG. Constraints which reduce the classical theory to a Bianchi I model have been imposed in the quantum theory as operator equations and their kernel has been computed. 
In the case of the most simple quantum states, the evolution coincides with a polymer quantisation of the corresponding minisuperspace model. 
The especially attractive feature of our model is its simplicity, being within the full theory while purely build on an Abelian gauge group. Issues like singularity resolution and the influence of the dynamics on coarse graining can thus be discussed explicitly, or transferred to the full theory directly from existing LQC calculations. Future work should also explore inhomogeneous perturbations of the Bianchi I background, which our model naturally incorporates, and study in particular the corrections they induce for the LQC dynamics. \\

{\bf Acknowledgements:}
This work was supported by a Feodor Lynen Research Fellowship of the Alexander von Humboldt-Foundation. Discussions with Emanuele Alesci, Jerzy Lewandowski, Daniele Oriti, Lorenzo Sindoni, J{\k{e}}drzej \'Swie\.zewski, Edward Wilson-Ewing, and Antonia Zipfel are gratefully acknowledged.

\end{document}